\documentclass[sigconf]{acmart}
\usepackage[utf8]{inputenc}

\usepackage{soul}
\usepackage{graphicx}
\usepackage{amsmath}
\usepackage{amssymb}
\usepackage{xspace}
\usepackage{cleveref}

\newcommand{\ie}{\emph{i.e.,}\xspace}
\newcommand{\eg}{\emph{e.g.,}\xspace}

\newcommand{\toolname}{\emph{Sosed}\xspace}

\newcommand{\secpart}[1]{\smallskip\textbf{#1.}\xspace}

\title{\toolname: a tool for finding similar software projects}

\begin{document}

\author{Egor Bogomolov}
\affiliation{JetBrains Research}
\email{egor.bogomolov@jetbrains.com}

\author{Yaroslav Golubev}
\affiliation{JetBrains Research}
\email{yaroslav.golubev@jetbrains.com}

\author{Artyom Lobanov}
\affiliation{JetBrains Research}
\email{artem.lobanov@jetbrains.com}

\author{Vladimir Kovalenko}
\affiliation{JetBrains Research, JetBrains N.V.}
\email{vladimir.kovalenko@jetbrains.com}

\author{Timofey Bryksin}
\affiliation{JetBrains Research}
\email{timofey.bryksin@jetbrains.com}

\begin{abstract}
    In this paper, we present \toolname, a tool for discovering 
    similar software projects. We use fastText to compute the embeddings of sub-tokens into a dense space for 120,000 GitHub repositories in 200 languages. Then, we cluster embeddings to identify groups of semantically similar sub-tokens that reflect topics in source code.
    We use a dataset of 9 million GitHub projects as a reference search base. To identify similar projects, we compare the distributions of clusters among their sub-tokens.
    The tool receives an arbitrary project as input, extracts sub-tokens in 16 most popular programming languages, computes cluster distribution, and finds projects with the closest distribution in the search base.
    We labeled sub-token clusters with short descriptions to enable \toolname to produce interpretable output.
    
    \toolname is available at \textit{\url{https://github.com/JetBrains-Research/sosed/}}. The tool demo is available at \textit{\url{https://www.youtube.com/watch?v=LYLkztCGRt8}}. The multi-language extractor of sub-tokens is available separately at \textit{\url{https://github.com/JetBrains-Research/buckwheat/}}.

\end{abstract}

\maketitle

\section{Introduction}\label{sec:introduction}

Identification of similar projects in a large set of open-source repositories can help in several software engineering tasks: rapid prototyping, program understanding, plagiarism detection~\cite{mens2014}. Additionally, it requires the development of new approaches to understand the meaning behind code and represent software projects at a large scale. In turn, if the developed methods can detect similar projects, they might be also applied in other software engineering tasks.

While popular search engines provide an option to search for web pages or images similar to the input, there is no common approach for finding similar software projects. For instance, prior work on similar projects detection leveraged several sources of data: Java API calls~\cite{mcmillan2012}, contents of README files~\cite{zhang2017}, user reactions in the form of GitHub stars~\cite{zhang2017}, tags on SourceForge~\cite{thung2012}.

Recently, several papers proposed to split code tokens into sub-tokens to improve results in method name prediction~\cite{AlonCode2Seq}, variable misuse identification~\cite{Hellendoorn2020Global}, and source code topic modeling~\cite{markovtsev2017topic}.
Following these advances, we suggest a novel approach to represent arbitrary fragments of code based on sub-token embeddings, \eg numerical representations in a dense space. We train sub-token embeddings with fastText~\cite{fasttext}, an algorithm for training word embeddings that takes into account both words and their subparts.

As prior work demonstrated, words with similar embeddings tend to be semantically related~\cite{schnabel2015}. We retrieve groups of related sub-tokens by clustering their embeddings with the spherical K-means algorithm~\cite{hornik2012}, a modification of the regular K-means~\cite{lloyd1982} that works with cosine distance. These clusters represent topics that occurred in a large corpus of source code. We represent code as a distribution of clusters among its sub-tokens.

We implemented the suggested approach to represent code as a tool for detecting similar projects called \toolname. We define similarity of projects as the similarity of the corresponding cluster distributions. To measure it, we suggest using either KL-divergence~\cite{kullback1951}, or cosine similarity of the distribution vectors.

\toolname identifies similar projects based solely on their codebase and supports 16 most popular languages. It does not make use of collaboration data (\eg GitHub stars) to avoid popularity bias. Currently, \toolname supports the search of similar repositories across 9 million repositories that comprise all unique public projects on GitHub as of the end of 2016. In future, we plan to update the dataset to use an up-to-date snapshot of Github.

An important feature of \toolname is the explainability of its output. We manually labeled the sub-token clusters with short descriptions of their topics. For each query result, we can provide descriptions of topics that contributed the most to the similarity measure.

The main contribution of our work is \toolname --- an open-source tool for finding similar repositories based on the novel code representation. \toolname provides explainable output, supports 16 programming languages, and searches across millions of reference projects.

The tool is available on GitHub~\cite{JbrSimilarRepositories}.
The part of \toolname used for sub-token extraction and language identification is also available as a standalone tool~\cite{JbrTokenizer}.

\section{Background}\label{sec:background}

Previous work on detecting similar repositories leveraged several sources of data. 
McMillan et al.~\cite{mcmillan2012} suggested CLAN, a Java-specific approach that detected similar Java applications by analyzing their API calls. The authors applied Latent Semantic Indexing~\cite{Deerwester1990IndexingBL} to an occurrence matrix, where columns represent projects, and rows represent API calls. The authors obtained vector representations of Java applications and defined the similarity of two projects as the cosine similarity of the corresponding vectors.

Aside from analyzing the code, several approaches to similarity search used data specific to code hosting platforms (\eg SourceForge~\cite{thung2012} or GitHub~\cite{zhang2017}).
Thung et al.~\cite{thung2012} used  the SourceForge's tags system to define similarity of the projects. Tags are short descriptions of project characteristics: category, language, user interface, and so on. Since some tags are more descriptive than others, the authors proposed to assign a weight to each tag. Then, they computed similarity of two projects from their sets of tags and their intersection.
Zhang et al.~\cite{zhang2017} measured similarity of projects hosted on GitHub based on the stars given by the same user in a short period of time and contents of the projects' README files.

The problem of detecting similar applications is also actively researched in the domain of mobile apps~\cite{chen2015, linares2016, li2017android, gonzalez2014}. The main difference from open source software projects is the data associated with each app. For apps in app stores, source code is often not openly accessible, but there are multiple other kinds of data available: description, images, permissions, user reviews, download size.

Another method related to measuring similarity of projects is topic modeling on code. The goal of topic modeling is to automatically detect topics in a corpus of unlabeled data, \eg software projects. The output of a topic modeling algorithm is a set of topics, and a distribution of topics in each item from the corpus. A topic is usually represented by a group of reference words or labels that are most frequent across data comprising the topic. 
According to the survey by Sun et al.~\cite{sun2016}, the most popular approach to topic modeling in software engineering is LDA~\cite{lda2003}. It treats source code as a bag of tokens, such as variable names, function names, and other identifiers. Markovtsev et al.~\cite{markovtsev2017topic} used ARTM~\cite{Vorontsov2015}, an algorithm similar to LDA, to identify topics across 9 million GitHub projects, which makes it, to the best of our knowledge, the largest study of topic modeling on source code.

\section{Description of the tool}\label{sec:internals}

In this work, we present \toolname, a tool for finding similar software projects based on a novel representation of code.

\begin{figure}[htp]
\centering
    \includegraphics[width=\columnwidth]{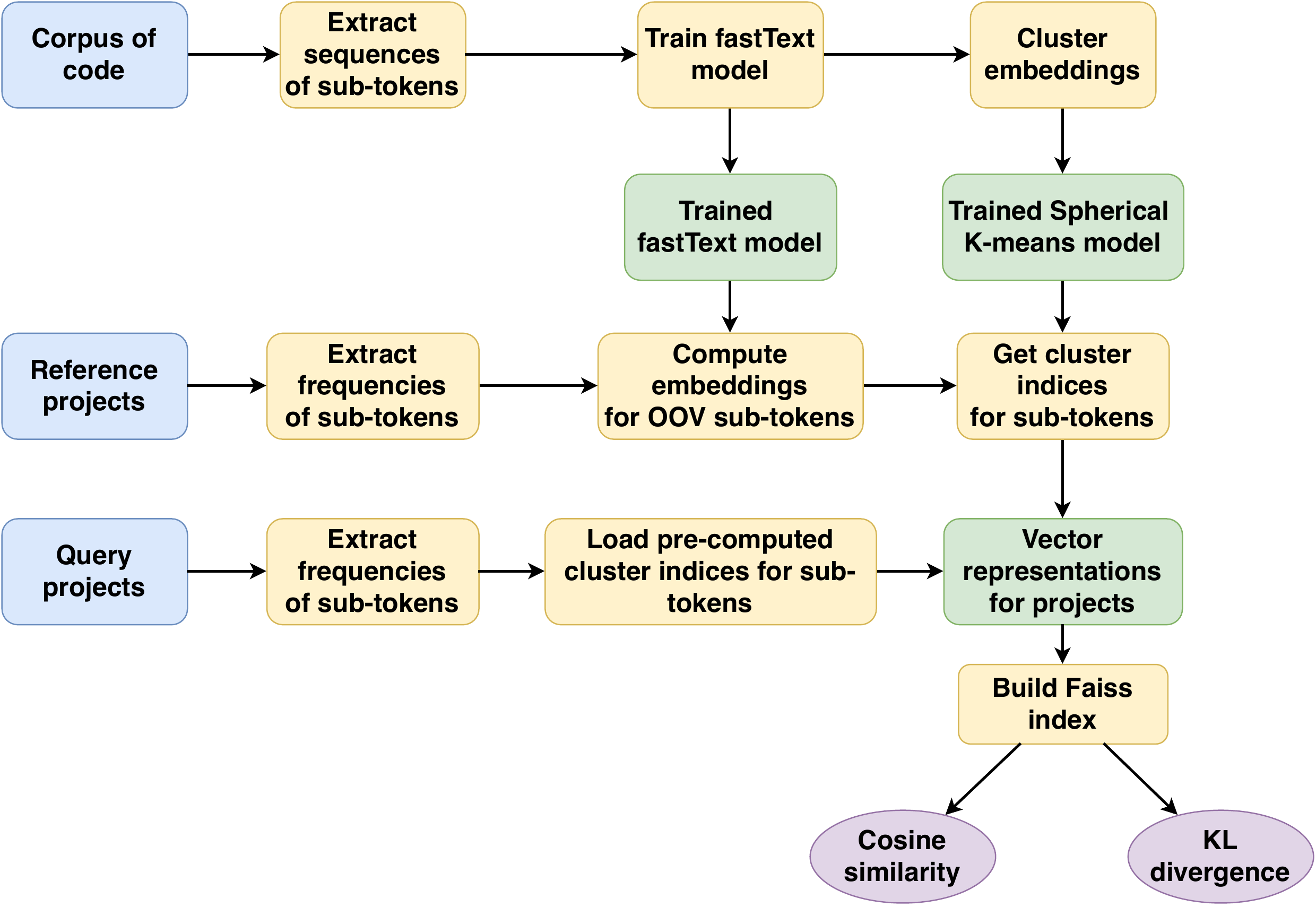}
    \centering
    \caption{Overview of the algorithm to compute projects' similarity}
    \label{fig:tool-overview}
\end{figure}

\secpart{Outline of \toolname's internals}
\Cref{fig:tool-overview} provides an overview of \toolname's internals. To find similar projects, we should define a search space, represent projects in a way suitable for searching, and set up a similarity measure.

As for the search space, we use the dataset of 9 million GitHub repositories collected by Markovtsev et al.~\cite{markovtsev2017topic}. To the best of our knowledge, it is the largest deduplicated dataset of software projects, which is suitable for our task straight-away.

As a preprocessing step, we transform projects into numerical vectors. Firstly, we train embeddings of sub-tokens on a large corpus of code~\cite{markovtsev2017dataworld} with fastText~\cite{fasttext}. Secondly, we find $K$ clusters of sub-tokens with spherical K-means algorithm~\cite{hornik2012}, where $K$ is a manually selected parameter. Finally, for each repository, we compute the distribution of clusters among its sub-tokens. The distribution for a project is a $K$-dimensional vector, where each component $C$ is a probability of cluster $C$ appearing among the project's sub-tokens. 

We implement two methods for measuring similarity of projects: explicitly computing KL-divergence~\cite{kullback1951} (\ie a measure of distribution similarity) of their cluster distributions, or computing cosine similarity of the distribution vectors. In both cases, we use Faiss~\cite{faiss} library to find the closest distributions.

In the rest of this section we describe parts of the tool in more details.

\secpart{Reference projects}\label{subsec:ref-projects}

For each repository, the dataset introduced by Markovtsev et al.~\cite{markovtsev2017topic} contains a set of all sub-tokens found in the project. We describe the process of extracting sub-tokens latter in this section.

The dataset is already cleared of both explicit and implicit forks (\ie copies of other projects that are not marked as forks on GitHub by its authors). It contains all the GitHub projects as of the end of 2016. Even though the projects in the dataset are not up-to-date, it allows us to implement the search in a vast amount of projects. In future, we plan to create an up-to-date version of the dataset.

\secpart{Training sub-token embeddings}\label{subsec:subtoken-embeddings}
For training sub-token embeddings, we use a dataset of identifiers extracted from 120,000 GitHub repositories~\cite{markovtsev2017dataworld}. It contains sequences of sub-tokens from files in approximately 200 programming languages.

We use fastText~\cite{fasttext} to compute embeddings of sub-tokens into a 100-dimensional space. Alongside with embeddings of input words, fastText also computes embeddings of encountered n-grams. It is helpful in the source code domain, because even at sub-token level there are some highly repetitive n-grams. Another important feature of fastText is its ability to compute embeddings for out-of-vocabulary (OOV) tokens: sub-tokens of reference projects not encountered in the corpus used for training embeddings. We computed embeddings for OOV sub-tokens with the trained fastText model, which gave us a set of 40 million known sub-tokens.

\secpart{Extracting sub-tokens from repositories}\label{subsec:subtokens}
A part of this work used for sub-token extraction and language identification might be useful for other tasks as well. To share it with the community and facilitate its reuse, we make it available as a separate project~\cite{JbrTokenizer}. The input of sub-token extractor is a list of either links to GitHub repositories or paths to local directories. The output is a list of all extracted sub-tokens and quantities of sub-tokens for each project.

On the first step of tokenization, we use \textit{enry}~\cite{enry} to recognize languages in files in each project. \textit{enry} is a Go-based language tool that employs several strategies to determine the language of a given file, including its name, extension, and content. \textit{enry} features the support of 382 languages, fast performance, and does not require a git repository to work, meaning that the input project can be any collection of files. 

When run on a directory, \textit{enry} outputs a JSON file with the recognized languages as keys and lists of files as values.
Using these keys, we filter languages that we are interested in. Based on the statistics on programming languages popularity~\cite{languages}, we currently support 16 languages, namely: C, C\#, C++, Go, Haskell, Java, JavaScript, Kotlin, PHP, Python, Ruby, Rust, Scala, Shell, Swift, and TypeScript. 

The next step of tokenization is extraction of identifiers. Since we are only interested in identifiers and names, we need to iterate over all the tokens in the file and gather only those that belong to specific types (excluding literals, comments, etc.). To do that, we employ two different tools. 12 out of 16 languages (including 10 most popular ones) are passed on to \textit{Tree-sitter}~\cite{treeSitter}, a fast parsing tool that uses language-specific grammars to parse a given file into an abstract syntax tree (AST).
We then filter the AST leaves to obtain various kinds of identifiers, names, constants, etc.

The four remaining languages (Scala, Swift, Kotlin, and Haskell) either do not have a \textit{Tree-sitter} grammar at the time of writing or the grammar is in development. The files in these languages are passed on to \textit{Pygments}~\cite{pygments} lexers. A \textit{Pygments} lexer splits the code into tokens, each of which also has a certain type. From the list of tokens, we extract those that are of interest to us: this includes the \texttt{token.Name} type by default, but for some languages it also makes sense to gather other types.

The last step of tokenization is splitting each token into sub-tokens. Following Markovtsev et al.~\cite{markovtsev2017topic}, we split the tokens by camel case and snake case, append short sub-tokens (less than three characters) to the adjacent longer ones, and stem sub-tokens longer than 6 characters using the Snowball stemmer~\cite{porter2001snowball}.

For a given project, we carry out identifier extraction and subtokenization for all files written in the supported languages and accumulate the results: in the end, the repository is represented as a dictionary with sub-tokens as keys and their counts as values.

\secpart{Clustering sub-token embeddings}\label{subsec:clustering}
We use the spherical K-means algorithm~\cite{hornik2012} to find clusters of similar sub-tokens. The algorithm is similar to the regular K-means~\cite{lloyd1982}, but it works with cosine distance instead of the Euclidean distance. Since we work with millions of high-dimensional vectors and cosine distance, other approaches like DBSCAN~\cite{dbscan1996} turn out to be too computationally expensive.

Spherical K-means requires choosing the number of clusters $K$ beforehand. We estimate an optimal number of clusters with gap statistic~\cite{gapStatistic}, a technique based on comparing the distribution of the inner-cluster distances with a uniform distribution. It has not shown any significant difference for the number of clusters above 256, so we decided to set $K$ to 256 to reduce the dimensionality of project representations at the next step.

Clusters represent groups of semantically similar sub-tokens. They can be seen as topics at the sub-token level. As in topic modeling, the topic can be guessed from a set of representatives. In our case, the representatives are the most frequent sub-tokens in the cluster and sub-tokens closest to the cluster center. To further elevate this information and make \toolname's output explainable, we manually labeled clusters with short descriptions by looking both at the representatives and projects where they are frequently used.

\secpart{Project representations}\label{subsec:project-vectors}
From the previous step, we get a mapping from sub-tokens to clusters. Then, we compute the distribution of clusters among sub-tokens in each project. For each repository, we get a $K$-dimensional vector where a coordinate along the dimension $C$ is equal to the probability of the cluster $C$ appearing among project's sub-tokens.

We applied the described technique to compute representations of 9 million repositories from the dataset of Markovtsev et al.~\cite{markovtsev2017topic}, which includes all unique projects (excluding both the explicit and implicit forks) on GitHub as of the end of 2016. This large set of projects forms the \toolname's search space.

\secpart{Searching for similar repositories}\label{subsec:search}
To find similar repositories to a given one, we should compute a cluster distribution for it. Firstly, we tokenize the project as previously described. Then, we collect pre-computed cluster indices for the sub-tokens encountered in reference projects. We do not compute embeddings for OOV sub-tokens in the new projects for two reasons. Firstly, their number is small, because the reference projects contain 40 million different sub-tokens. Secondly, OOV sub-tokens may refer to libraries and technologies that emerged after the reference dataset had been collected , \ie the end of 2016. In this case, the embeddings will not reflect the underlying semantics of sub-tokens.

We implement two methods to compare cluster distributions between projects from a query and reference projects: direct computation of KL-divergence~\cite{kullback1951} between two distributions and cosine similarity of the distribution vectors. Cosine similarity equals to the inner product of the normalized distribution vectors. KL-divergence can be expressed by the following formula:
$$
D_{KL}(P_Q || P_R) = \sum\limits_{c \in Clusters} P_Q(c)\log{\dfrac{P_Q(c)}{P_R(c)}}\\ ,
$$
\noindent where $P_Q$ and $P_R$ are cluster distributions for a query and a reference project, respectively.
Finding a reference project $R$ that minimizes KL-divergence for the given query project is equivalent to maximizing the following function:
$$
\sum\limits_{c \in Clusters} P_Q(c)\log{P_R(c)}.
$$

The function is an inner product of the cluster distribution $P_Q$ and a point-wise logarithm of the distribution $P_R$. Thus, both for KL-divergence and cosine similarity, the search of similar projects reduces to maximizing an inner product between two vectors. 

We utilized the Faiss~\cite{faiss} library to find vectors giving the maximal inner product. Faiss transforms reference vectors into an indexing structure that can be further used for querying. The indexing structure used in our work does not introduce a significant memory overhead, which allows us to use it with a large search space.

To enable the tool to provide explanations for project similarity, we find sub-token clusters corresponding to the terms that contributed the most to the vectors' inner product. Within the tool's output, we display their contributions alongside with manually given labels and sub-tokens from these clusters.

\section{Evaluation}

To the best of our knowledge, the only approach to evaluate the output of algorithms for finding similar projects used in previous work~\cite{mcmillan2012,thung2012,zhang2017} is conducting a survey of developers. 

Since \toolname works with programming projects in 16 languages, thorough evaluation of its performance without diving deep into specific ecosystems becomes challenging. We plan to conduct a survey of a large group of programmers with different expertise in order for its results to be reliable.

For now, we evaluated \toolname's output on a set of 94 GitHub projects that comprises top-starred repositories in different languages. The results are available on our GitHub page~\cite{JbrSimilarRepositories}. For example, top-5 most similar projects to TensorFlow\footnote{\url{https://github.com/tensorflow/tensorflow/}} are deep learning and machine learning frameworks. For Bitcoin\footnote{\url{https://github.com/bitcoin/bitcoin/}} \toolname detected other open-sourced cryptocurrencies. Among projects similar to Python\footnote{\url{https://github.com/python/cpython/}} we found Brython,\footnote{\url{https://github.com/brython-dev/brython/}} a Python implementation running in a browser. 

\section{Conclusion}\label{sec:conclusion}

Finding similar software projects among a large set of repositories might be beneficial for practical software engineering tasks like quick prototyping and program understanding. Aside from that, it requires development of new methods for representing source code, which can find application in other software-related tasks.

We created a novel approach to represent code based on the topic distribution among its sub-tokens. We implemented it as a tool for finding similar software repositories called \toolname. The main features of \toolname are explainability of its output, support of 16 programming languages, and independence of project popularity. \toolname is available on GitHub~\cite{JbrSimilarRepositories,JbrTokenizer}.

For now, \toolname searches among a set of 9 million GitHub projects. While it is a large set of data, open-source community grew rapidly over the recent years~\cite{GithubStats}. In order to catch up with the growth of the open-source ecosystem, we plan to collect a new dataset, which will contain an up-to-date set of GitHub projects.

Implementation of open-source tools for the novel ideas has several benefits. This way, we can quickly evaluate the method's performance, check its practical applicability, and gather feedback from the tool's users. We encourage others to create open-source software based on the developed methods in order to speed up communication and evolution in the research community.

\bibliographystyle{ACM-Reference-Format}
\bibliography{paper}

\end{document}